\documentclass{ws-ijmpcs}

\begin{document}

\markboth{JINGQING ZHANG} %%running head: author
{Observation of the $X(1840)$ at BESIII}%%running head: title

\catchline{}{}{}{}{}
%
%%%%%%%%%%%%%%%%%%%%%%%%%%%%%%%%%%%%%%%%%%%%%%%%%%%%%%%%%%%%%%%%%%%%

\title{OBSERVATION OF THE $X(1840)$ AT BESIII}

\author{JINGQING ZHANG\\
ON BEHALF OF THE BESIII COLLABORATION}

\address{Institute of High Energy Physics, Chinese Academy of Sciences\\
Beijing 100049, China\\
zhangjq@ihep.ac.cn}

%\author{SECOND AUTHOR}
%
%\address{Group, Laboratory, Address\\
%City, State ZIP/Zone, Country\\
%second\_author@domain\_name}

\maketitle

%\begin{history}
%\received{Day Month Year}
%\revised{Day Month Year}
%\end{history}

\begin{abstract}
Observation of the $X(1840)$ in the $3(\pi^+\pi^-)$ invariant mass in
$J/\psi\rightarrow\gamma 3(\pi^+\pi^-)$ at BESIII is reviewed.
With a sample of $225.3\times10^6$ $J/\psi$ events collected with the BESIII
detector at BEPCII, the $X(1840)$ is observed
with a statistical significance of $7.6\sigma$.
The mass, width and product branching fraction of the $X(1840)$ are determined.
The decay $\eta^\prime\rightarrow 3(\pi^+\pi^-)$ is searched for,
and the upper limit of the branching
fraction is set at the 90\% confidence level.
\end{abstract}

\ccode{PACS numbers: 13.20.Gd, 12.39.Mk}

\section{Introduction}	
Within the standard model framework, the strong interaction is described
by Quantum Chromodynamics (QCD), which suggests the existence
of the unconventional hadrons, such as glueballs, hybrid states and multiquark states.
The establishment of such states remains one of the main interests in experimental particle physics.

Decays of the $J/\psi$ particle are ideal for the study of the hadron spectroscopy
and the searching for the unconventional hadrons.
In the decays of the $J/\psi$ particle, several observations in the mass region 1.8 GeV/c$^2$ - 1.9 GeV/c$^2$
have been presented in different experiments\cite{Bai:2003sw}\cdash\cite{Ablikim:2011pu},
such as the $X(p\bar{p})$\cite{Bai:2003sw}\cdash\cite{BESIII:2011aa},
$X(1835)$\cite{Ablikim:2005um}$^{,}$\cite{Ablikim:2010au}, $X(1810)$\cite{Ablikim:2006dw}$^{,}$\cite{Ablikim:2012ft}
 and $X(1870)$\cite{Ablikim:2011pu}.

\section{Observation of the $X(1840)$}
Recently, %%%the paper Ref.~\refcite{Ablikim:2013spp} presented the observation of a new structure, the $X(1840)$.
using a sample of $225.3\times10^{6}$ $J/\psi$ events\cite{Ablikim:2012cn} collected
with BESIII detector\cite{Ablikim:2009aa} at BEPCII\cite{Bai:2001dw},
the decay of $J/\psi\rightarrow\gamma 3(\pi^+\pi^-)$ was analyzed\cite{Ablikim:2013spp}, and
the $X(1840)$ was observed in the $3(\pi^+\pi^-)$ mass spectrum
with a statistical significance of $7.6\sigma$.

\begin{figure}[pbht]
\centerline{\includegraphics[width=0.6\textwidth]{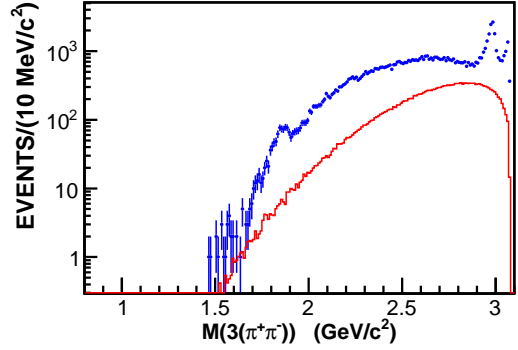}}
\vspace*{8pt}
\caption{Distribution of the invariant mass of $3(\pi^+\pi^-)$. The dots with error
bars are data; the histogram is phase space events with
an arbitrary normalization. \label{m6pi}}
\end{figure}

\begin{figure}[pbht]
\centerline{\includegraphics[width=0.6\textwidth]{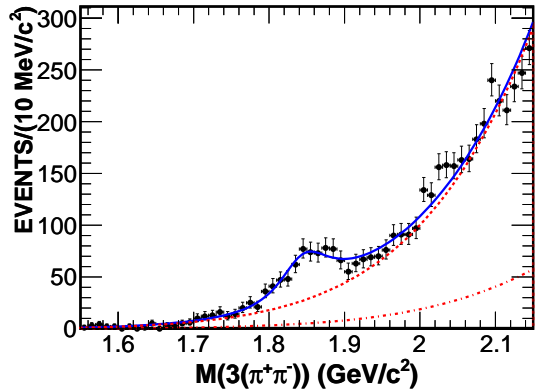}}
\vspace*{8pt}
\caption{The fit of mass spectrum of $3(\pi^+\pi^-)$. The dots with error bars are data;
the solid line is the fit result. The dashed line represents all the backgrounds, including
the background events from $J/\psi\rightarrow\pi^03(\pi^+pi^-)$ (dash-dotted line, fixed
in the fit) and a third-order polynomial representing other backgrounds. \label{m6pi_fit}}
\end{figure}

The $3(\pi^+\pi^-)$ invariant mass spectrum is shown in Fig.~\ref{m6pi},
where the $X(1840)$ can be clearly seen.
The parameters of the $X(1840)$ are extracted by an unbinned maximum likelihood fit.
In the fit, the background is described by two contributions:
the contribution from $J/\psi\rightarrow\pi^03(\pi^+\pi^-)$
and the contribution from other sources.
The contribution from $J/\psi\rightarrow\pi^03(\pi^+\pi^-)$ is determined from
MC simulation and fixed in the fit (shown by the dash-dotted line in Fig.~\ref{m6pi_fit}).
The other contribution is
described by a third-order polynomial.
The signal is described by a Breit-Wigner function
modified with the effects of the detection efficiency, the detector resolution, and the phase
space factor.
The fit result is shown in Fig.~\ref{m6pi_fit}.
The mass and width of the $X(1840)$ are $M=1842.2\pm4.2^{+7.1}_{-2.6}$ MeV/c$^2$
and $\Gamma=83\pm14\pm11$ MeV, respectively;
the product branching fraction of the $X(1840)$ is $\mathcal{B}(J/\psi\rightarrow\gamma X(1840))
\times\mathcal{B}(X(1840)\rightarrow 3(\pi^+\pi^-)) = (2.44\pm0.36^{+0.60}_{-0.74})\times10^{-5}$.
In these results, the first errors are statistical and the second errors are systematic.

\begin{figure}[pbht]
\centerline{
\includegraphics[width=0.67\textwidth]{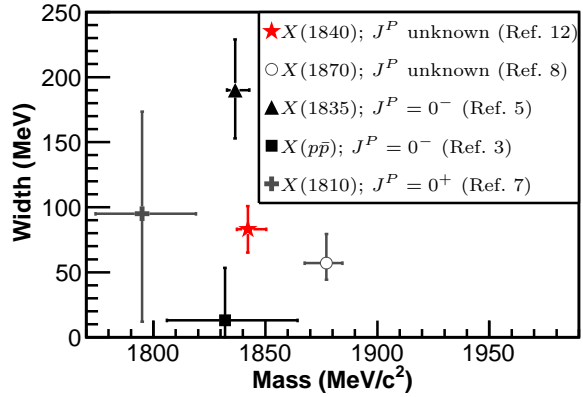}
\put(-137, 138) {\scriptsize{$X(1840)$; $J^{P}$ unknown (Ref.~\refcite{Ablikim:2013spp})}}
\put(-137, 123.5) {\scriptsize{$X(1870)$; $J^{P}$ unknown (Ref.~\refcite{Ablikim:2011pu})}}
\put(-137, 109) {\scriptsize{$X(1835)$; $J^{P} = 0^-$ (Ref.~\refcite{Ablikim:2010au})}}
\put(-137, 94.5) {\scriptsize{$X(p\bar{p})$; $J^{P} = 0^-$ (Ref.~\refcite{BESIII:2011aa})}}
\put(-137, 80) {\scriptsize{$X(1810)$; $J^{P} = 0^+$ (Ref.~\refcite{Ablikim:2012ft})}}
}
\vspace*{8pt}
\caption{Comparisons of observations at BESIII. The error bars
include statistical, systematic, and, where applicable, model uncertainties. \label{comp_mw}}
\end{figure}

Figure~\ref{comp_mw} shows the comparisons of the $X(1840)$ with other
observations at BESIII\cite{Ablikim:2013spp}.
The comparisons indicate that at present one can not distinguish whether the $X(1840)$ is a new state or the
signal of a $3(\pi^+\pi^-)$ decay mode of an existing state.

\section{Search for the decay of $\eta^\prime\rightarrow 3(\pi^+\pi^-)$}
With the same data sample, the decay of $\eta^\prime\rightarrow 3(\pi^+\pi^-)$
was searched for\cite{Ablikim:2013spp}.
The mass spectrum of the $3(\pi^+\pi^-)$ is shown in Fig.~\ref{m6pi}, where
no events are observed in the $\eta^\prime$ mass region.
With the Feldman-Cousins frequentist approach\cite{Feldman:1997qc}, the upper limit of
the branching fraction is set to be
$\mathcal{B}(\eta^\prime\rightarrow 3(\pi^+\pi^-)) < 3.1\times10^{-5}$
at the 90\% confidence level, where the systematic uncertainty is taken into account.

\section{Summary}
With a sample of $225.3\times10^{6}$ $J/\psi$ events collected at BESIII,
the decay of $J/\psi\rightarrow\gamma3(\pi^+\pi^-)$ was analyzed\cite{Ablikim:2013spp}.
The $X(1840)$ was observed in the $3(\pi^+\pi^-)$ invariant mass spectrum. The mass, width
and product branching fraction of the $X(1840)$ are
$M=1842.2\pm4.2^{+7.1}_{-2.6}$ MeV/c$^2$, $\Gamma=83\pm14\pm11$ MeV
and $\mathcal{B}(J/\psi\rightarrow\gamma X(1840))
\times\mathcal{B}(X(1840)\rightarrow 3(\pi^+\pi^-)) = (2.44\pm0.36^{+0.60}_{-0.74})\times10^{-5}$,
respectively. The decay $\eta^\prime\rightarrow3(\pi^+\pi^-)$ was searched for.
No events were observed in the $\eta^\prime$ mass region
and the upper limit of the branching fraction was set to be
$\mathcal{B}(\eta^\prime\rightarrow 3(\pi^+\pi^-)) < 3.1\times10^{-5}$
at the 90\% confidence level.

%%\section*{Acknowledgments}
%%
%%This section should come before the References. Dedications and funding
%%information may also be included here.
%%

%\begin{thebibliography}{000} %for 3 digits

\end{document}